# The phenomenon of resonance in the continuous phase transition of finite-size systems: A passage from Classical World to Quantum World through the resonance?

Yiannis F. Contoyiannis[1,*], Stelios M. Potirakis[1]

1. Department of Electrical and Electronics Engineering, University of West Attica, 12244 Aigaleo-Athens, Greece.

* yiaconto@uniwa.gr

**Abstract**

In finite-size systems undergoing a continuous phase transition, the passage from the symmetric phase to the broken-symmetry phase is accomplished through a hysteresis zone, up to spontaneous symmetry breaking (SSB). In the present work, we find that a resonance phenomenon takes place within this zone. This resonance is manifested as a maximization of the mean waiting time as a function of temperature inside the hysteresis region. An interesting issue concerns how this resonance is connected with the existence of particles (tachyons) or quasiparticles (kink solitons) within the hysteresis zone. Finally, we introduce the idea that this resonance delineates a continuous passage from a “classical” phase to a “quantum” phase for a binary system, such as the three-dimensional Ising model, which belongs to the same universality class as a fermion–antifermion system or, more generally, a matter–antimatter system.

## 1. Introduction

In a recent work [1], we have shown that, in second-order phase transitions of finite-size systems, a temperature interval emerges. This hysteresis zone is delimited by the pseudocritical temperature and the temperature at which spontaneous symmetry breaking (SSB) occurs. In the present work, we aim to understand the behavior of the critical exponent that governs the dynamics of thermal fluctuations as the temperature decreases within the hysteresis zone. In other words, we investigate how the critical character of the system (in particular, the 3D Ising model) is degraded inside the hysteresis region. This question constituted the original motivation for the present study.

During the investigation of this issue, we identified the existence of a resonance phenomenon between the mean values of the laminar lengths (waiting times), extracted from the time series of magnetization fluctuations, and the corresponding temperatures within the hysteresis zone. In particle physics, resonances may decay into particles or quasiparticles. Within the hysteresis zone, we observe the presence of tachyons (as particles) and kink solitons (as quasiparticles), which possess imaginary mass; consequently, these excitations “survive” in Euclidean space [1]. Finally, this resonance appears to separate two distinct phases: a classical phase and a quantum thermal fluctuation phase.

## 2. Theoretical Background

### 2.1 Elements of second-order phase transitions according to critical $\phi^4$ theory

For thermal systems of infinite-size, the established theoretical description of spontaneous symmetry breaking (SSB) in second-order phase transitions is formulated in terms of the Ginzburg–Landau (G–L) free energy, truncated at fourth order in the order parameter (the $\phi^4$ theory). Specifically, the G–L free energy $U(\phi)$ is given by [2,3]:

$$U(\phi) = a \cdot \phi^4 + b \cdot \phi^4, \qquad (1)$$

Where $\phi$ denotes the order parameter. The coefficients in Eq. (1) satisfy the conditions $a > 0, b > 0$ in the symmetric phase, i.e., when the temperature (the control parameter) is equal to the critical temperature ($T = T_c$; green curve in Fig. 1(a)). In contrast, $a < 0, b > 0$ in the symmetry-broken phase, corresponding to $T < T_c$ (blue and red curves in Fig. 1(b) and Fig. 1(c), respectively).

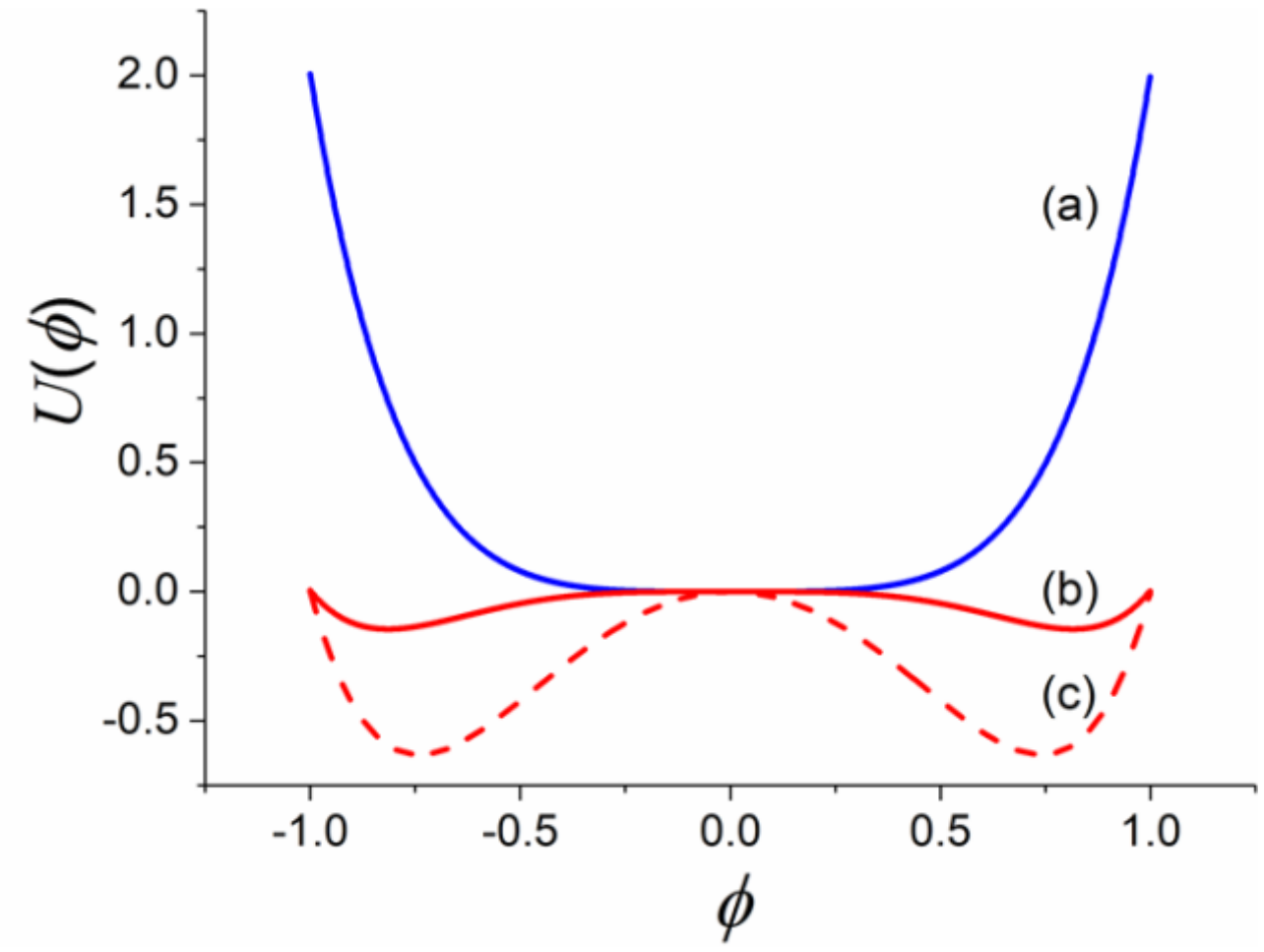

**Fig. 1.** The G-L free energy $U(\phi)$ vs. the order parameter $\phi$ for the second-order transition of a thermal system of infinite size: (a) symmetrical phase (achieved for $T = T_c$). (b) SSB (for an arbitrarily selected $T < T_c$). (c) SSB (for another arbitrarily selected $T < T_c$).

As illustrated in Fig. 1, in the symmetric phase the G–L free energy exhibits a single minimum at a specific value of the order parameter. When the temperature decreases below the critical temperature, a degenerate set of minima appears, leading to spontaneous symmetry breaking. For an infinite-size system, a key feature is the absence of communication between the degenerate free-energy minima, which are disconnected in configuration space. Consequently, once equilibrium is established for $T < T_c$, the system must select one specific free-energy minimum [2], and SSB persists for all temperatures below $T_c$.

As the control parameter (e.g., the temperature in a thermal system) crosses below its critical value, the unstable critical point at $\phi = 0$ ceases to exist, and stable vacuum states emerge (red curves in Fig. 1). According to the standard theory, the system relaxes into one of these two vacua, which constitutes the manifestation of SSB. A classical system cannot simultaneously occupy both vacua after SSB, since this would imply the quantum mechanical principle of superposition of states, which is not applicable to classical phase transitions.

## 2.2 The Z(N) spin systems, 3D-Ising model

For a Z(N) spin system, spin variables are defined as: $s(a_j) = e^{i2\pi a_j/N}$ (lattice vertices $j = 1, 2, \dots j_{max}$), with $a_j = 0, 1, 2, \dots N-1$. For the case Z=2 there are two possible states of spin which are +1, -1, for the case of Z=3, there are three possible states of spin: $\left|-\frac{2\pi}{3}\right\rangle$, $|0\rangle$, and $\left|+\frac{2\pi}{3}\right\rangle$, and so on. Such a thermal system can be studied within the Landau approach for the description of critical phenomena [4], using the notion of G-L free energy in the general can be described as a polynomial function $U(\phi)$ of the order parameter $\phi$:

$$U(\phi) = \frac{1}{2} r_0 \phi^2 + \frac{1}{4} u_0 \phi^4 + \frac{1}{6} c_0 \phi^6 + \frac{1}{8} s_0 \phi^8 + \cdots. \tag{2}$$

It is noted that, in the symmetric phase, all the coefficients in Eq. (2) are positive. This equation can provide a description of the first- and second-order phase transitions, which can be achieved by utilizing terms up to $\phi^6$ or up to $\phi^4$, respectively. Therefore, the symmetric phase, in second order phase transition is described by Eq. (1).

The 3D-Ising model is a lattice model of ferromagnetism in Statistical Mechanics [4]. It successfully describes the continuous phase transition in equilibrium as well as more specialized topics as SSB of $\phi^4$ theory [2-4].

According to the theory of critical phenomena, natural systems are classified into universality classes, which are characterized by the values of the so-called critical exponents. The two-dimensional Ising model was solved analytically by Onsager, and its critical exponents are therefore known exactly. In contrast, the three-dimensional Ising model has not yet been solved analytically and is accessible only through numerical and approximate methods. Its critical behavior has been extensively investigated using renormalization group techniques, Monte Carlo simulations, the conformal bootstrap, and related approaches [5–8].

An efficient algorithm for generating equilibrium configurations of Ising models is the Metropolis algorithm. In this algorithm, configurations at fixed temperature are sampled according to the Boltzmann statistical weight $e^{-\beta H}$, where $H$ is the Hamiltonian of the spin system. For nearest-neighbor interactions, the Hamiltonian can be written as:

$$H = -\sum_{\langle i,j \rangle} J_{ij} s_i s_j, \tag{3}$$

where $s_i = \pm 1$ are the spin variables and $J_{ij}$ denotes the interaction strength.

It is well known [2–4] that this model undergoes a second-order phase transition when the temperature falls below a critical value. For a finite cubic lattice of size $20^3$ in three dimensions (the 3D Ising model), the critical temperature—more precisely, the pseudocritical temperature due to finite-size effects—has been determined to be $T_{pc} = 4.545\ (J_{ij} = 1)$ [9].

A single sweep of the entire lattice defines the algorithmic unit of time. In our numerical experiments performed using the Metropolis algorithm, we record the mean magnetization $M$, which plays the role of the order parameter. The trajectory generated by the simulation constitutes a time series of fluctuations of the order parameter. A time series of length $N_{\text{iter}}$ can be represented in discrete form, with the time step corresponding to one lattice sweep. In the present study, the simulations were performed for $N_{iter} = 200{,}000$ sweeps.

In Fig. 2(a), we present the distribution of magnetization values at the pseudocritical temperature $T_{pc} = 4.545$ for a $20^3$ lattice. To avoid artifacts associated with histogram bin widths, we also analyze the system dynamics using a phase-space representation. Accordingly, Fig. 2(b) shows a segment of the phase-space trajectory of the order parameter $M$.

In statistical mechanics, a phase-space diagram is traditionally defined as a plot of momentum $p = m\frac{\mathrm{d}x}{\mathrm{d}t}$ versus position $x$. This concept can be generalized by replacing the position variable with a field variable; correspondingly, the momentum is given by the temporal derivative of the field. By convention, the mass is set to $m = 1$.

In this generalized phase-space representation, the horizontal axis corresponds to the values of the magnetization field $M$, while the vertical axis represents its temporal derivative.

In the discrete-time formulation relevant for numerical simulations, the phase-space points are given by $\left(\frac{M_{n+1}-M_n}{\Delta\tau}\right)$, where the time interval between successive measurements is $\Delta\tau = (n+1) - n = 1$, corresponding to one lattice sweep.

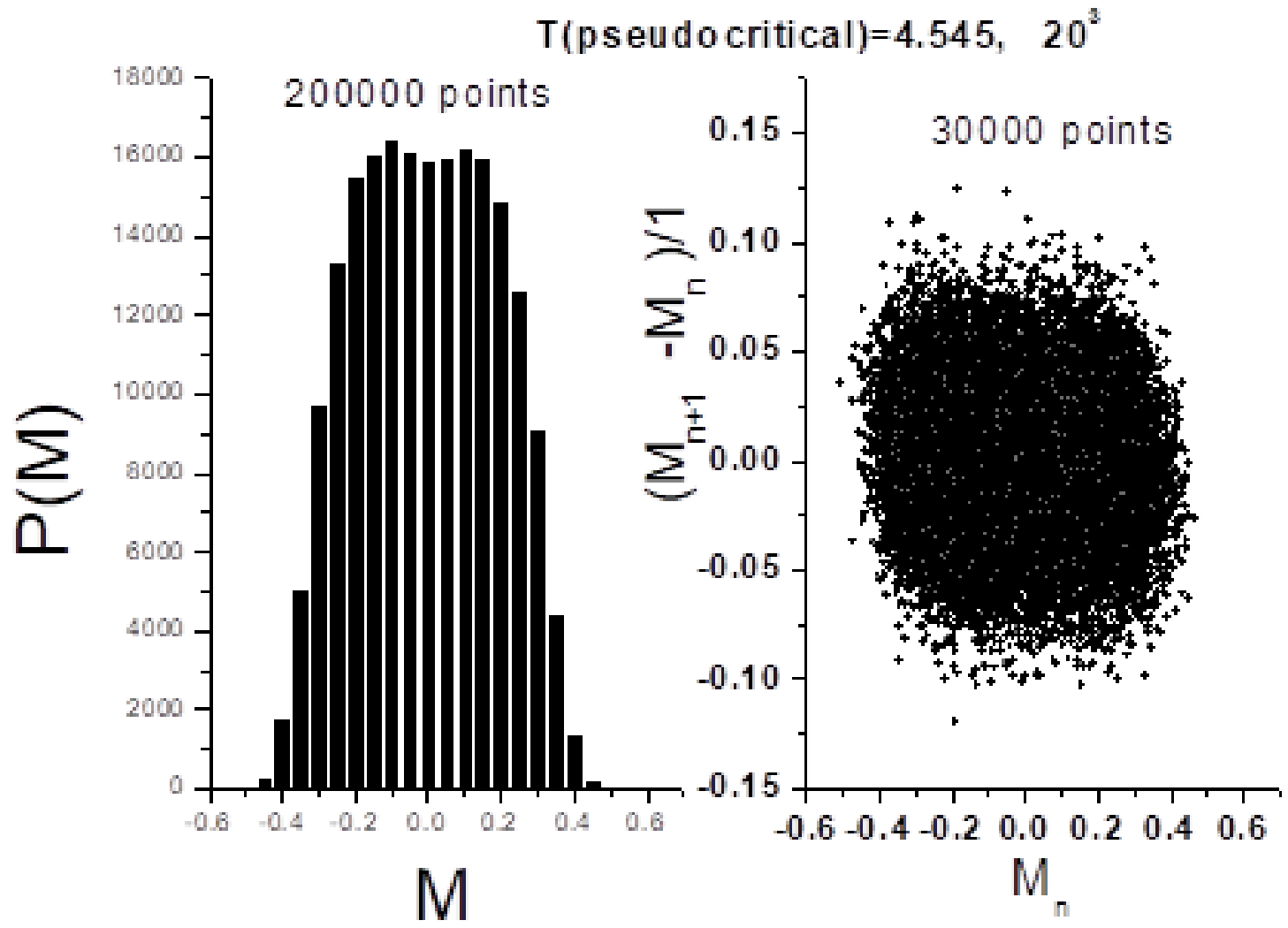

**Fig. 2.** (a) Distribution of order-parameter fluctuations at the pseudocritical temperature $T = T_{pc} = 4.545$, obtained using the Metropolis algorithm. (b) A segment consisting of 30,000 points of the phase-space diagram at the pseudocritical temperature.

### 2.3 The dynamics of critical state.

The dynamics of the fluctuations of the order parameter at the critical state can be determined analytically for a large class of complex systems introducing the so-called critical map, which can be approximated as [9]:

$$\phi_{n+1} = \phi_n + u \cdot \phi_n^z + \varepsilon_n, \tag{4}$$

where $\phi_n$ is the $n$th sample of the scaled order parameter, $u > 0$ is a coupling parameter, $z$ stands for a characteristic exponent, associated with the critical isothermal exponent $\delta$ as:

$$z = \delta + 1 \quad (5)$$

and $\varepsilon_n$ stands for the non-universal stochastic noise necessary for the creation of ergodicity [10]. Equation (4) describes the type-I intermittency which is a physical mechanism very important in many phenomena [11]. Intermittency consists of successive time intervals where low fluctuations occur –called "laminar lengths"– interrupted, in a chaotic way, by temporal intervals of high fluctuations –called "bursts". It is noted that the laminar lengths in the case of a time series $\phi$, correspond to the waiting times, i.e., the number of consecutive time series values lie within the time series values' interval $[\phi_0, \phi_L]$ (called "laminar region"), bounded by the fixed-point $\phi_0$ and a number of different values within the $\phi$ values range, which are called "ends of laminar regions" and denoted as $\phi_L$. A very important quantity that can be calculated in the Ising models, like in almost all dynamical systems in real- or algorithmic-time systems, is the distribution of waiting times. The basic information which can be exported from such a distribution is the quantitative information about the dynamics of a system. In the cases where this distribution is a scale free function, such as a power-law, the signature of critical state of the system could be revealed. Therefore, the correct calculation of waiting times is of great importance. As shown in [9], the fluctuations of the order parameter at the critical point follow the temporal dynamics of type I intermittency, for which it is known [11] that the distribution of appropriately defined waiting times (in other words, the distribution of laminar lengths), follows a power-law of the form:

$$P(L) \sim L^{-p}. \quad (6)$$

As shown in [9], the exponent $p$ is related to the isothermal critical exponent with the relation:

$$p = 1 + 1/\delta. \quad (7)$$

As it is known [4] that $\delta > 1$. So, from Eq. (7) we have that the critical state exists when $p \in [1,2)$. In [9], the value of the exponent p for critical 3D-Ising model was estimated in very good agreement with the predicted theory of critical phenomena. Specifically, by applying the method of critical fluctuations (MCF), which we have introduced for this purpose, it was found that $p = 1.21$ for critical 3D-ising model [9]. Exactly on the critical point, the characteristic exponent is the isothermal critical exponent $\delta$. For the 3D-Ising model this exponent has the value $\delta = 4.8$ [12]. So, from Eq. (7) we obtain that the theoretical value of exponent $p$, is $p = 1.208$. Therefore, the estimated value ($p = 1.21$) is perfect agreement with the theoretical value.

### 3. The Hysteresis zone before the SSB in finite-size systems

It has been found that in any finite-size thermal physical system exhibiting a second-order phase transition according to the $\phi^4$ theory, a temperature zone emerges below the critical temperature that could be considered as pseudocritical in finite-size size systems [1]. An example of such a system is the classic paradigm of the 3D-Ising spin system.

Inside this zone, the second-order phase transition critical point remains, even though the mathematical symmetry in the Landau free energy is broken (Fig. 1). This region between the critical (in fact pseudocritical) temperature $T_c$ and the SSB temperature $T_{SSB}$ ($T_{SSB}$<$T_{pc}$), where the critical point appears for the last time, is called the hysteresis zone and its properties have been studied in detail in [13]. In Fig. 3 we present the distribution of the magnetization values of the 3D-Ising model in a lattice $20^3$ as it is described in Section 2.2. The distributions are presented as the temperature inside zone drops. The characteristic in Fig. 3 is the appearance of the two lobes which communicated each to other. We start immediately after the critical temperature whose distribution we presented in Fig. 2 and we reach the temperature where the two lobes have almost separated but still, for the last time, coexist. The temperature at which this occurs is the temperature at where SSB occurs. As has been shown in [1,13], the width $\Delta T = T_{pc} - T_{SSB}$ is depended from the length of the system through a power-law of the form $\Delta T \sim d^{-\mu}$ ($\mu > 1$), where $d$ is the length of lattice. Therefore, the smaller the system size, the wider this zone is. Consequently, for an infinite-size system ($d \to \infty$) the zone $\Delta T$ vanishes $\Delta T = 0$. Therefore, according to the above-presented finite-size approach, is in perfect agreement with the aforementioned established theory, i.e., $d \to \infty$ for the SSB, in terms of G-L $\phi^4$ theory for infinite-size systems. After the SSB, only one lobe appears according to the Subsection 2.1, either for the positive or the negative magnetization, inside the one vacuum of Fig. 1. Within the zone $\Delta T$ three fixed points appear. These are the initial unstable point (0,0), which acts as a repellor, and the two symmetric attractors, which evolve to the stable vacua of Fig. 1 when the SSB is completed. Due to the fact that the initial unstable point (repellor) remains as this transition evolves, the zone is characterized as hysteresis zone.

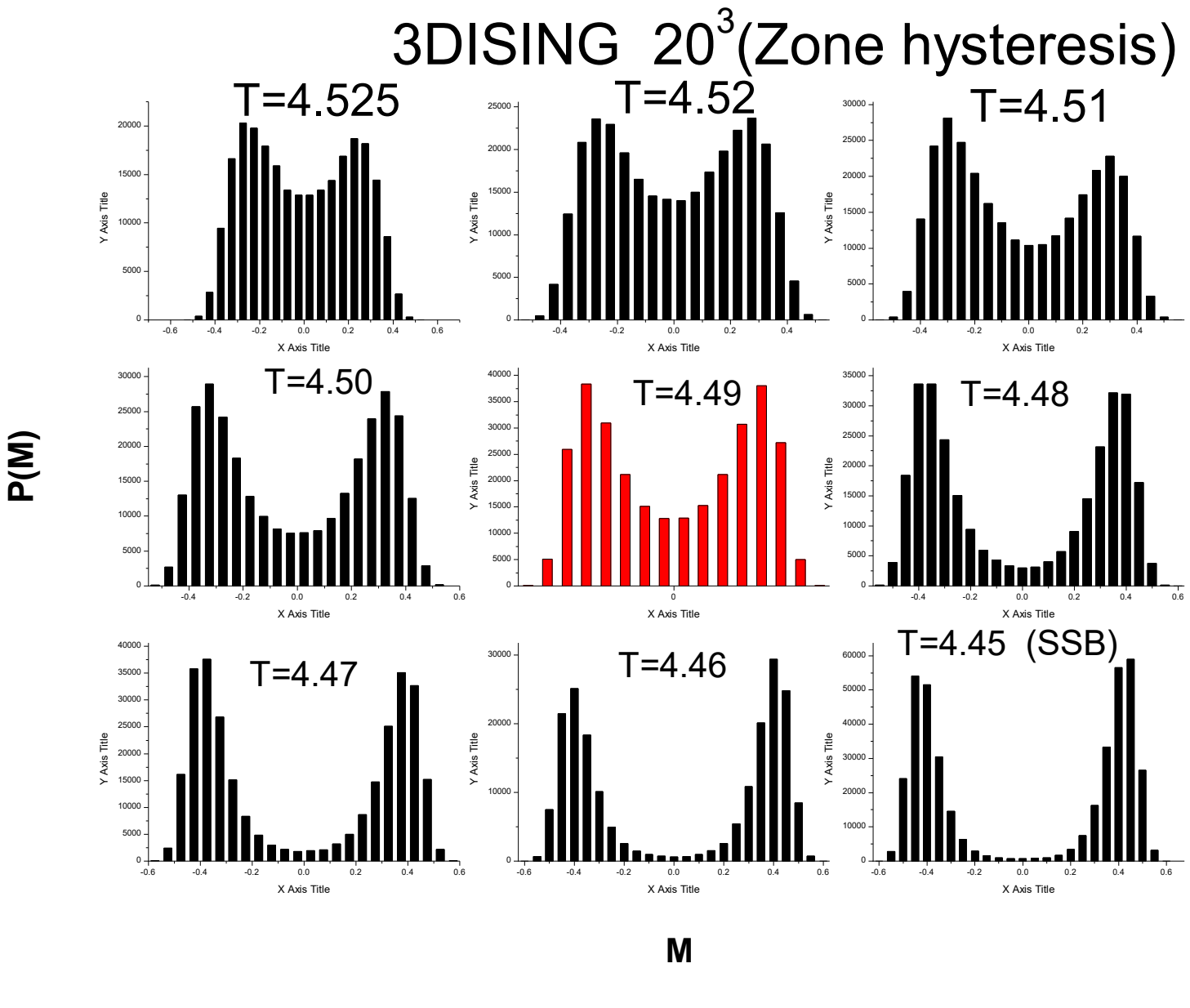

**Fig. 3.** Nine distributions are presented between the temperatures $T = 4.525$ up to the SSB temperature, $T_{SSB} = 4.45$, in the numerical model of 3D-Ising in a $20^3$ lattice. The distribution with the red color is particularly important as will be explained below. Distributions for temperatures from $T = 4.525$ to $T = 4.46$ correspond to 300,000 points long magnetization time series. For $T_{SSB} = 4.45$ the distribution corresponds to a 400,000

points long time series (larger statistics to better show the symmetry of the final separation of the lobes).

In Fig. 4 we also present the phase space, for the SSB temperature (corresponding the $T_{SSB} = 4.45$ distribution of Fig. 3), to clearly demonstrate the separation of the two lobes.

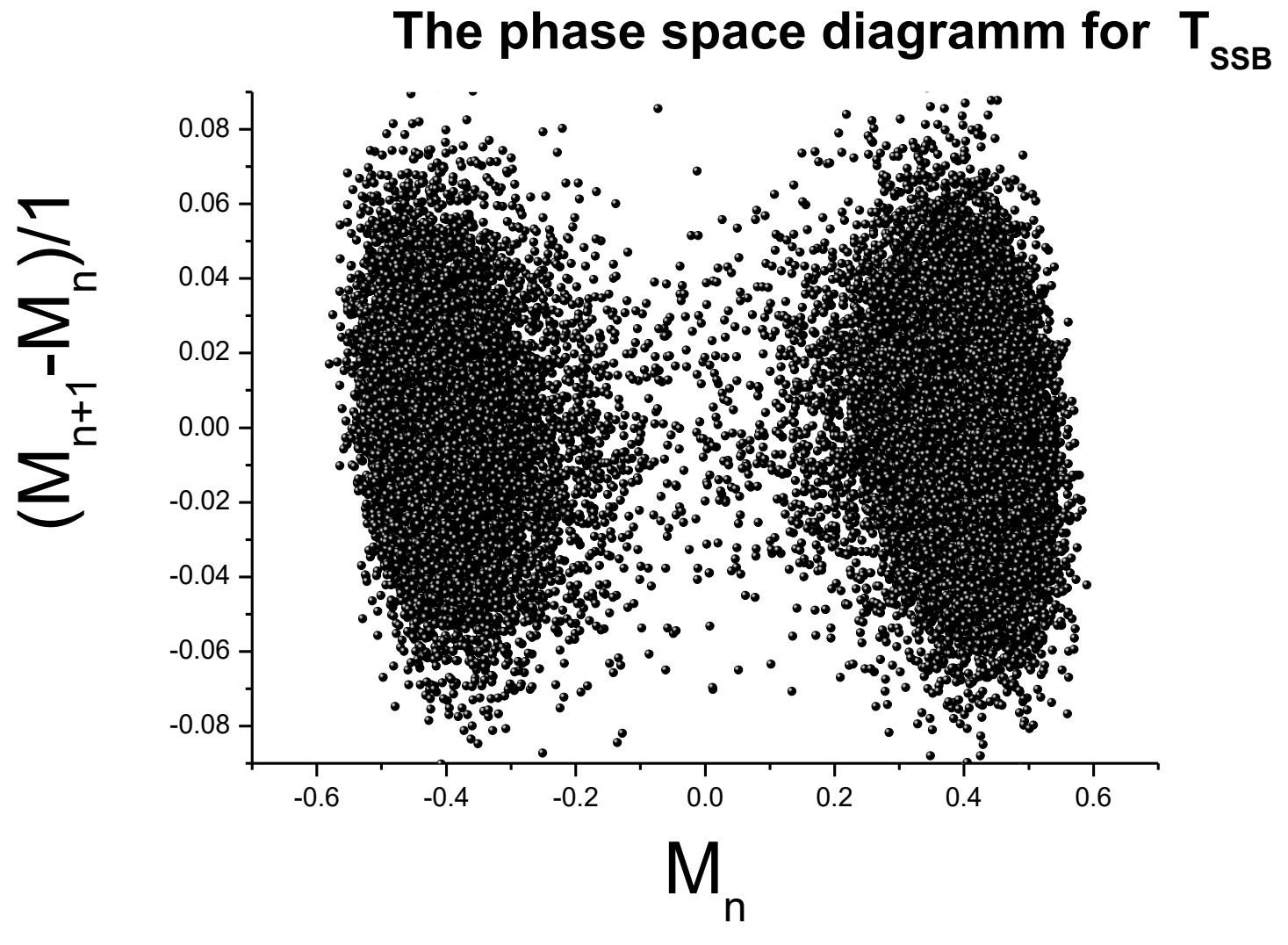

**Fig. 4.** Phase space diagram of the 3D-Ising numerical model $T_{SSB} = 4.45$, in a $20^3$ lattice.

**4. Criticality inside the hysteresis zone: The calculation methods.**

As mentioned in Section 2.2, the dynamics of the order parameter (magnetization) fluctuations the obeys the critical intermittency (Eq. (4)) and the laminar length distribution follows a power law with exponent $p = 1.21$ both at the theoretical level and as well as numerically. In this section we will check each one of the nine temperatures of the SSB zone presented in Fig. 3 for the existence of critical dynamics, and for each case that will be found to exhibit critical dynamics we will calculate the exponent $p$. To check for the existence of critical dynamics in a time series, as already mentioned (see Section 2.3), the method of critical fluctuations (MCF) [9,14] is used. To apply the MCF one has to calculate the distribution of laminar lengths. Here we will demonstrate the application of MCF to some of the temperatures of Fig. 3, for example for $T = 4.52$, and the same procedure is valid for the other temperatures too.

In the application of MCF, two characteristic values of the order parameter are used. These values are the fixed-point, $\phi_0$, which in one-dimensional iterative maps like the map described by Eq. (4) is determined according to the turning point method [15,16], and the "end of laminar region", $\phi_L$ (in positive or in negative values due of symmetry of the distributions), which is a varying parameter. These values are marked on Fig. 5(a) as red and blue horizontal lines, respectively. Thus, the laminar lengths are estimated inside the laminar regions $[\phi_0, \phi_L]$, i.e.:

$$\phi_0 < \phi < \phi_L. \quad (8)$$

The position of the fixed point $\phi_0$ for a perfectly symmetric distribution of magnetization values is zero, i.e., $\phi_0 = 0$, because in the critical state the order parameter, here the average magnetization, $M$, must be zero. The perfectly symmetric state occurs for $N_{iter} \to \infty$. The distributions of Fig. 3 have emerged, as already mentioned, using time series of $300{,}000$ or $400{,}000$ points which is a satisfactory number and is closer to a physical reality than infinity. So, one can perform the calculations on positive or negative values since the obtained results are very close as has been found in similar studies [17]. In the following we will use the symbol of magnetization $M$ in the position of the symbol $\phi$. The position of the end laminar $\phi_L$ is found by shifting the blue line, which defines the end of the laminar region, until it results to a laminar lengths distribution that is closer to the power law, if of course something like this exists.

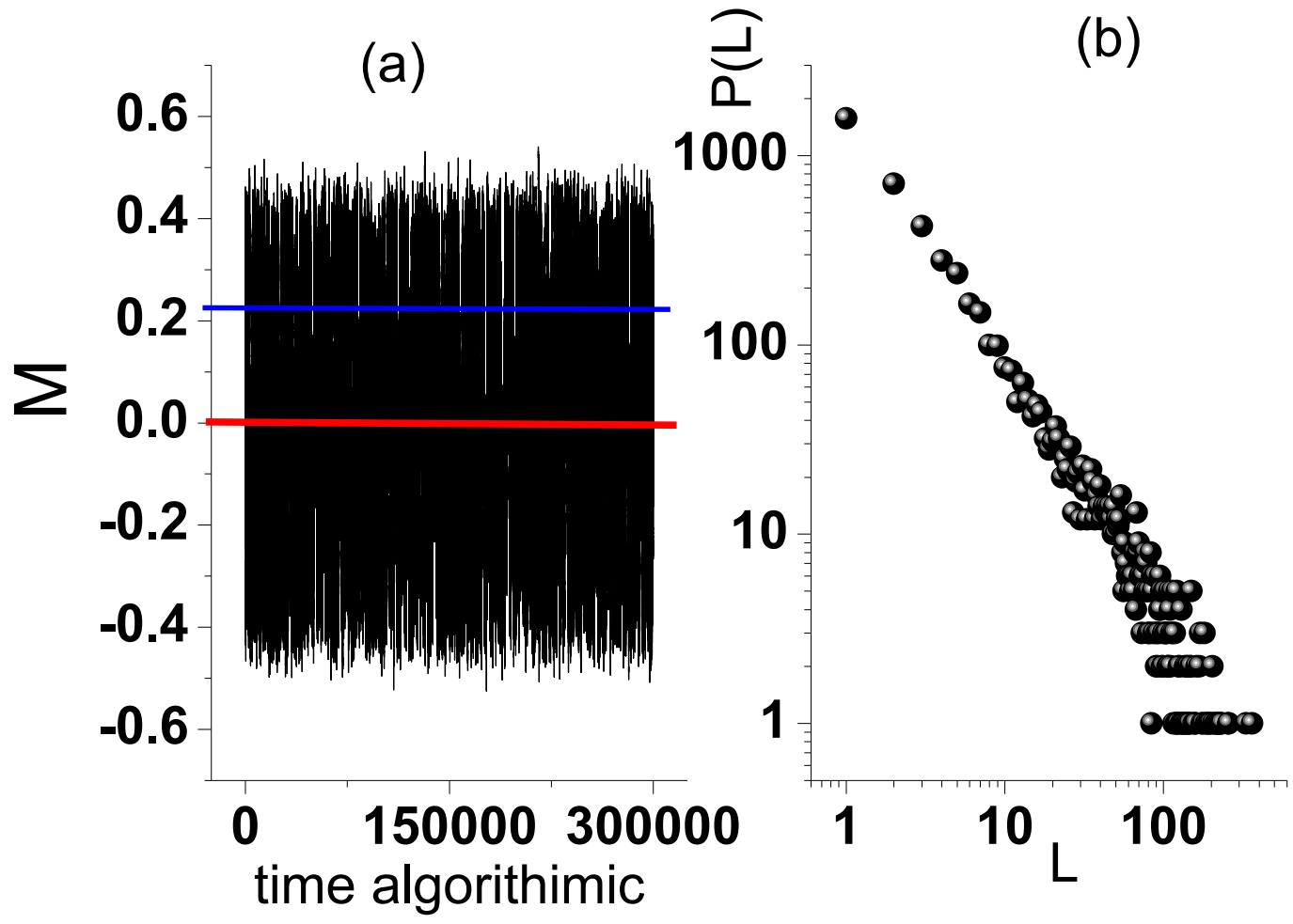

**Fig. 5.** The magnetization time-series for $T = 4.52$ (inside the hysteresis zone). The horizontal red line denotes the fixed point $M_0$ and the blue line denote the end, $M_L$, of the laminar region, respectively, used for the application of MCF according to Eq. (8) (see text) (b) Distribution of the laminar lengths inside the laminar region [$M_0 = 0$, $M_L = 0.22$].

In the distribution of the laminar lengths of Fig. 5b we see that the tail of the distribution (beyond the first 30 points) curves and does not follow the linearity of the first points in the log-log plot. So, for the first 30 points one can use typical fitting methods (least squares) to calculate the exponent $p$ but with such fitting methods, one cannot calculate what exponents the tail gives because the deviations of the points in the tail are much larger in relation to the deviations of the first points, and therefore it cannot result to a credible fit. So, in such cases one loses any information that the tail ($L \gg 1$) carries and keeps only the first points. But almost all the distributions of the waiting times, and much more in the case of intermittency that we examine here, the power laws are valid only for $L \gg 1$ [11,18,19,20]. So, for the exact estimation of the exponents one has to use a new method based on wavelet analysis, in which the calculations are made for the $L \gg 1$. This method has recently been introduced as a wavelet-based detection of scaling behavior in noisy experimental data [21]. Thus, long scales carry important information and cannot be excluded from the analysis. The wavelet basis used to develop the specific algorithm is the Haar wavelet. The wavelet basis elements are denoted: $\psi_{j,k}(x) = 2^{\frac{j}{2}}\psi(2^j x - k)$, with $j = \cdots, -1, 0, 1, \ldots$ and $k = -j, \ldots, -1, 0, 1, \ldots, j$, whereas the wavelet analysis coefficients are denoted $d_{j,k}$. For $j = k = 0$ one gets the coarse-graining description of the analysis. In the framework of this description, the coefficients of the analysis can and do ignore the noise of the signal to be analyzed [21]. This behavior has been used to develop an algorithm that applies to any discrete distribution, $P(i)$, $i = 1, 2, \ldots, N_P$, associated with real or numerical time series. This algorithm can and does answer the question of whether the analyzed distribution embeds a power-law behavior, how close or far is it from the power-law, and calculates the corresponding power-law exponent $p$. In other words, it provides a kind of "fitting" that can be applied to laminar lengths' distribution $P(L)$, without carrying the pathogeny of standard curve fitting-based methods due to noise in the experimental data, especially at the high values of the laminar lengths where the poor statistics of the points in the tails of the distributions lead to results of questionable reliability. The mother function of the Haar wavelets is defined, using the Heaviside step function, $\Theta(z)$, as:

$$\psi_H(x) = \Theta\left(\frac{\Delta}{2} - x\right)\Theta(x) - \Theta\left(x - \frac{\Delta}{2}\right)\Theta(\Delta - x), \quad (9)$$

for $x \in (0, \Delta]$, while the wavelet analysis coefficients for the expansion of a power-law function (appropriate for a finite-size system) $F(x) = \begin{cases} 0, & x \notin [\Delta_{min}, \Delta_{max}) \\ cx^{-p}, & \Delta_{min} \le x < \Delta_{max} \end{cases}$ can be written as [21]:

$$d_{j,k} = c\sqrt{\frac{2^j}{\Delta}}\left(\sum_{i=max\left(\left[k\frac{\Delta}{2^j}\right],1\right)}^{\left[k\frac{\Delta}{2^j}+\frac{\Delta}{2^{j+1}}\right]} i^{-p} - \sum_{i=\left[k\frac{\Delta}{2^j}+\frac{\Delta}{2^{j+1}}\right]+1}^{\left[(k+1)\frac{\Delta}{2^j}\right]} i^{-p}\right), \quad (10)$$

where $c$ is a normalization constant and $[z]$ stands for the integer part of a variable $z$.

One can then define the following quantities [21] for the discrete case:

$$\lambda = \frac{\frac{d_{00}}{d_{10}}}{\frac{d_{10}}{d_{20}}} = \frac{d_{00}d_{20}}{d_{10}^2} = \frac{\left(\sum_{i=1}^{\left[\frac{\Delta}{2}\right]} P(i) - \sum_{\frac{\Delta}{2}}^{\Delta} P(i)\right)\left(\sum_{\mathrm{i}=1}^{\left[\frac{\Delta}{8}\right]} P(i) - \sum_{\left[\frac{\Delta}{8}\right]}^{\left[\frac{\Delta}{4}\right]} P(i)\right)}{\left(\sum_{\mathrm{i}=1}^{\left[\frac{\Delta}{4}\right]} P(i) - \sum_{\left[\frac{\Delta}{4}\right]}^{\left[\frac{\Delta}{2}\right]} P(i)\right)^2} \quad (11)$$

and

$$R = \frac{d_{00}}{d_{10}} = \frac{1}{\sqrt{2}}\frac{\left(\sum_{i=1}^{\left[\frac{\Delta}{2}\right]} P(i) - \sum_{\frac{\Delta}{2}}^{\Delta} P(i)\right)}{\left(\sum_{\mathrm{i}=1}^{\left[\frac{\Delta}{4}\right]} P(\mathrm{i}) - \sum_{\left[\frac{\Delta}{4}\right]}^{\left[\frac{\Delta}{2}\right]} P(i)\right)}, \quad (12)$$

with $8 < \Delta < N_P$.

The proposed method has the following steps:

1. We apply Eq. (11) to calculate $\lambda$ as a function of $\Delta$ up to a value $\Delta_{max} \le N_P$. We plot $\lambda$ vs $\Delta$, and since we are interested in the convergence of $\lambda$ [21], we focus on the right-most part of the plot (for large $\Delta$), while up to the last 10 points, i.e., $n = \Delta_{max} - 9, \dots, \Delta_{max}$ [21] are enough to deduce a conclusion. The value of $\Delta_{max}$ is determined in the tail of the distribution of laminar lengths according to the next criterion.

2. quantify the previous step by calculating the distance of $\lambda$ from the value $\lambda = 1$, which denotes an exact power-law, by calculating the quantity:

$$Q_{\lambda} = \frac{1}{10} \sum_{n=\Delta_{max}-9}^{\Delta_{max}} (1 - \lambda_n)^2. \tag{13}$$

The closer the $Q_{\lambda}$ is to the value 0 the closer to the power-law is the distribution. So, the position of $\Delta_{max}$ is located where the quantity $Q_{\lambda}$ comes closest to zero.

2. Then, we produce the $R$ vs $\Delta$ plot, using Eq. (12). From the convergence region of the plot, i.e., for $n = \Delta_{max} - 9, \dots, \Delta_{max}$ (see Eq. (13)), a mean value, $\langle R \rangle$, is obtained for the quantity $R$.

3. We consider the test function $f(i) = c_f i^{-p}, i = 1,2,..\Delta_{max}$, where $c_f$ is a constant factor, and we solve Eq. (12) numerically for the given $\langle R \rangle$ value with respect to $p$, to estimate the value of $p$ exponent which corresponds to $\langle R \rangle$.

The power-law test function mentioned in Step 4 is used only when $Q_{\lambda} < order(10^{-3})$, which indicates an exact power-law, and further if $p \in [1,2)$ this indicates that the underlying system is in critical state. In case that $order(10^{-3}) < Q_{\lambda} < order(10^{-2})$, this indicates that the examined laminar lengths' distribution is close to power-law. In this case, in Step 4 we can use a truncated power-law test function $g(i) = c_g i^{-p} e^{-iq}, i = 1,2,..\Delta_{max}$, where $c_g$ is a constant factor, instead of a power-law test function $f(i)$ to further investigate. So, we quantify a power-law with an exponential correction, and if the exponents are found to be $p \in [1,2)$ and $q \approx 0$, this indicates that the underlying system is in critical state. In cases that $Q_{\lambda}$ turns out to be higher than the $order(10^{-2})$ there is no power-law distribution.

If one applies the above algorithm for the case of $T = 4.52$, will find that from Eqs. (11), (13) the smallest value of the quantity $Q_{\lambda}$ is $Q_{\lambda} = 2.49 \cdot 10^{-5}$ when the laminar region is [0. 0.22]. In Fig. 6a we present the 10 last values of the quantity $\lambda$ which are all on the line $\lambda = 1$. This means that the distribution of laminar lengths (Fig. 5b) is an exact power low. For the determination of exponent $p$ we find from Eq. (12) the last 10 values of parameter $R$ with mean field $\langle R \rangle = 0.7964$. In Fig. 6b we present the 10 last values of the quantity $R$ which are all on the line $0.7964$. Finally, we proceed in the fourth step solving the numerical equation for this mean value $\langle R \rangle$ we find that the exponent $p = 1.21 \in [1,2)$, i.e., the time series at $T = 4.52$ is in critical state.

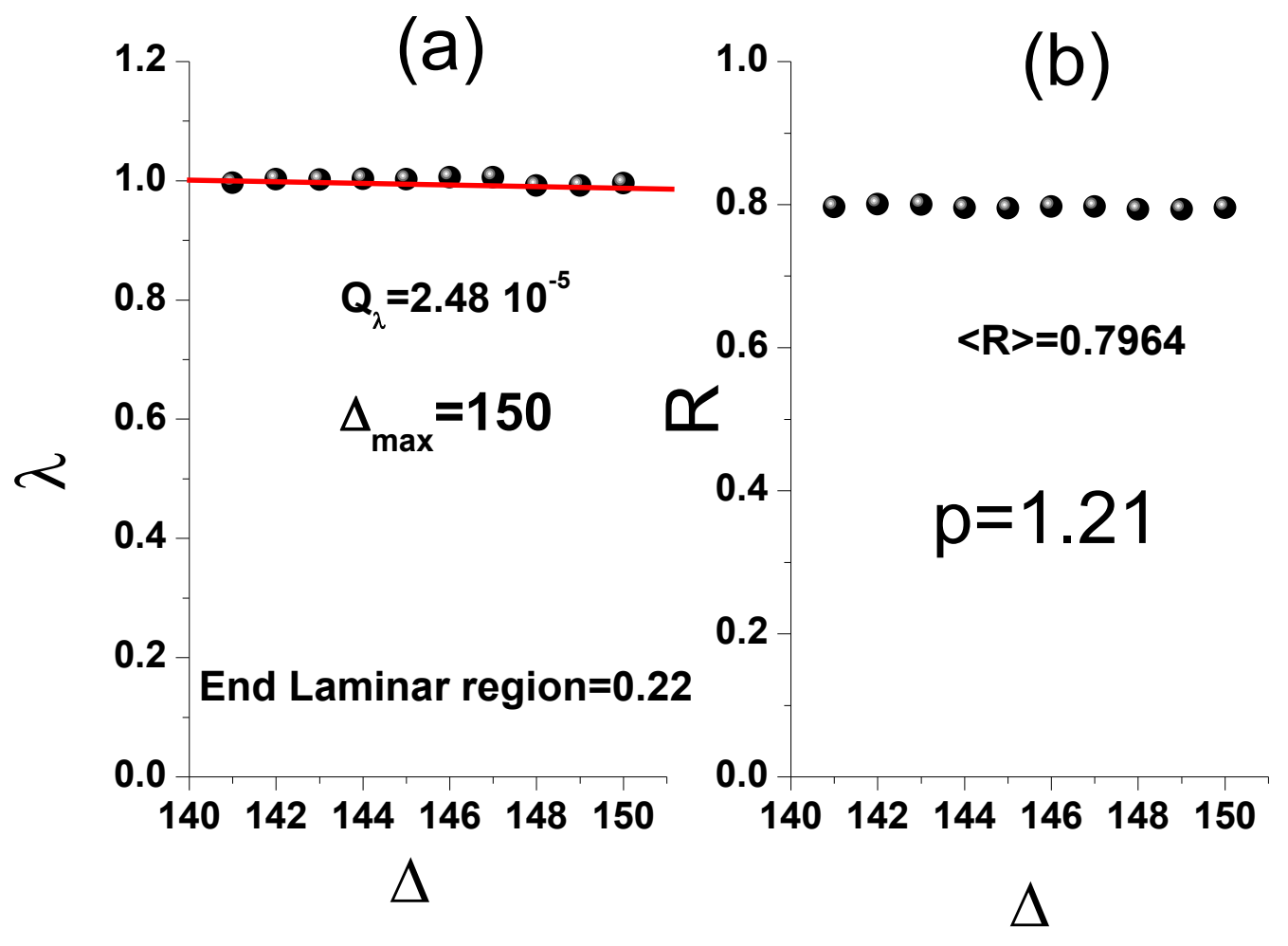

**Fig. 6.** (a) Plot of $\lambda$ vs. $\Delta$ with $\Delta_{max} = 150$. The horizontal red line corresponds to the criterion $\lambda = 1$, which denotes an exact power-law distribution. (b) Plot of $R$ vs. $\Delta$ with $\Delta_{max} = 150$. The mean value of these 10 values is $\langle R \rangle = 0.7964$ and the power-law exponent was finally estimated to be $p = 1.22 \in [1,2)$, indicating critical dynamics.

## 5. Results and Comments

Following the analysis procedure described in Section 4 and applied in Fig. 6a, we calculate the corresponding quantities for the remaining eight temperatures inside the hysteresis zone. The complete set of results is presented in Fig. 7.

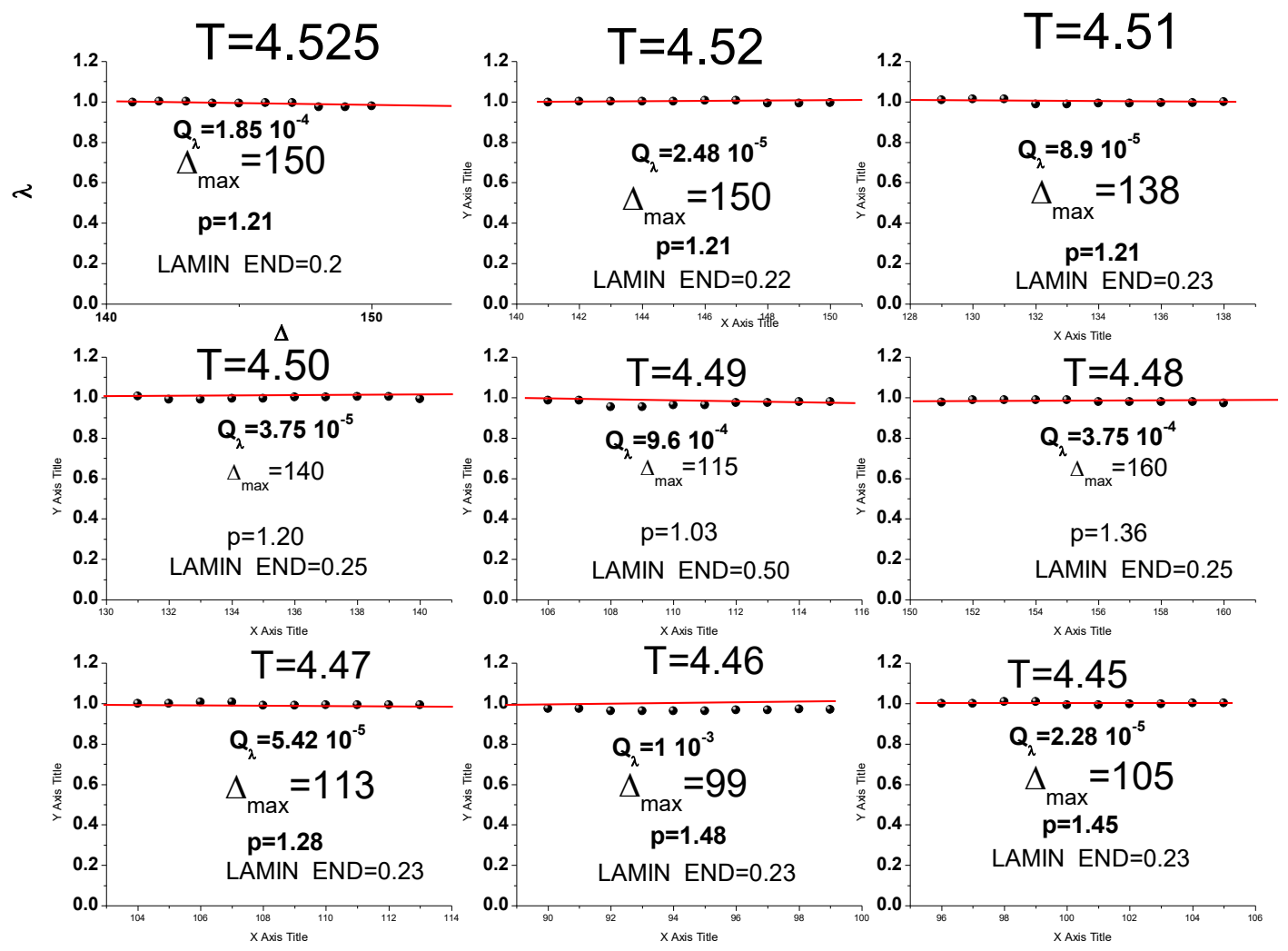

Fig. 7. Criticality analysis based on the Method of Critical Fluctuations (MCF) and wavelet-based detection of scaling behavior for the temperatures 4.525, 4.51, 4.52, 4.50, 4.49, 4.48, 4.47, 4.46, 4.45.

The above results show that:

- For the temperatures 4.525, 4.52, 4.51, and 4.50, the exponents of the distribution of laminar lengths are 1.21, 1.21, 1.21, and 1.20, respectively. Consequently, we recover the same exponent as that of the critical state. Thus, in this part of the hysteresis zone, a complete degeneration of criticality occurs at successive temperatures. The obtained exponent is very close to the theoretical value $p = 1.208$, corresponding to the universality class of the 3D Ising model.
- For the temperatures 4.48, 4.47, 4.46, and 4.45, the exponents of the laminar-length distributions are 1.30, 1.28, 1.48, and 1.45, respectively. All these values lie within the critical exponents' interval $1 \leq p < 2$. Therefore, in this part of the hysteresis zone, the critical dynamics associated with the initial critical fixed point persist. Moreover, as shown in [22], a power-law distribution of waiting times with exponent $p \in [1,1.5)$ indicates that critical organization is governed by a dominant Lévy process. In this regime, long laminar lengths survive. For $p \in [1.5,2)$, a gradual transition from Lévy to Gaussian statistics occurs, implying truncation of very long laminar lengths.

Thus, at all the above-mentioned temperatures inside the hysteresis zone, perfect criticality is preserved. The temperature $T = 4.49$, which lies approximately at the center of the hysteresis interval, is excluded here and discussed separately, as it exhibits a special behavior.

## 6. The Phenomenon of Resonance inside the Hysteresis Zone

From Fig. 7 we observe that for $T = 4.49$ the exponent takes the value $p = 1.03$. This value lies very close to the lower bound of the critical interval $p \in [1,2)$. Furthermore, the laminar region at $T = 4.49$ exhibits the largest width compared to all other temperatures. As shown in Fig. 8, the laminar interval extends over $[0,0.5]$.

The laminar region corresponding to the best power-law fit is marked in Fig. 8 between the red and blue arrows. The blue arrow indicates that the end of the laminar region coincides with the maximum of the lobe, i.e., it lies within the emerging stable vacuum of the theory (see Fig. 1). This corresponds to the symmetry-broken phase, where criticality ceases to exist.

At this temperature, a competition arises between the unstable initial critical point and the stable vacuum state. This competition limits the critical dynamics, yielding an exponent $p$ close to unity. Thus, the temperature $T = 4.49$ acts as a boundary separating a regime of degenerate criticality from the pathway toward complete symmetry breaking.

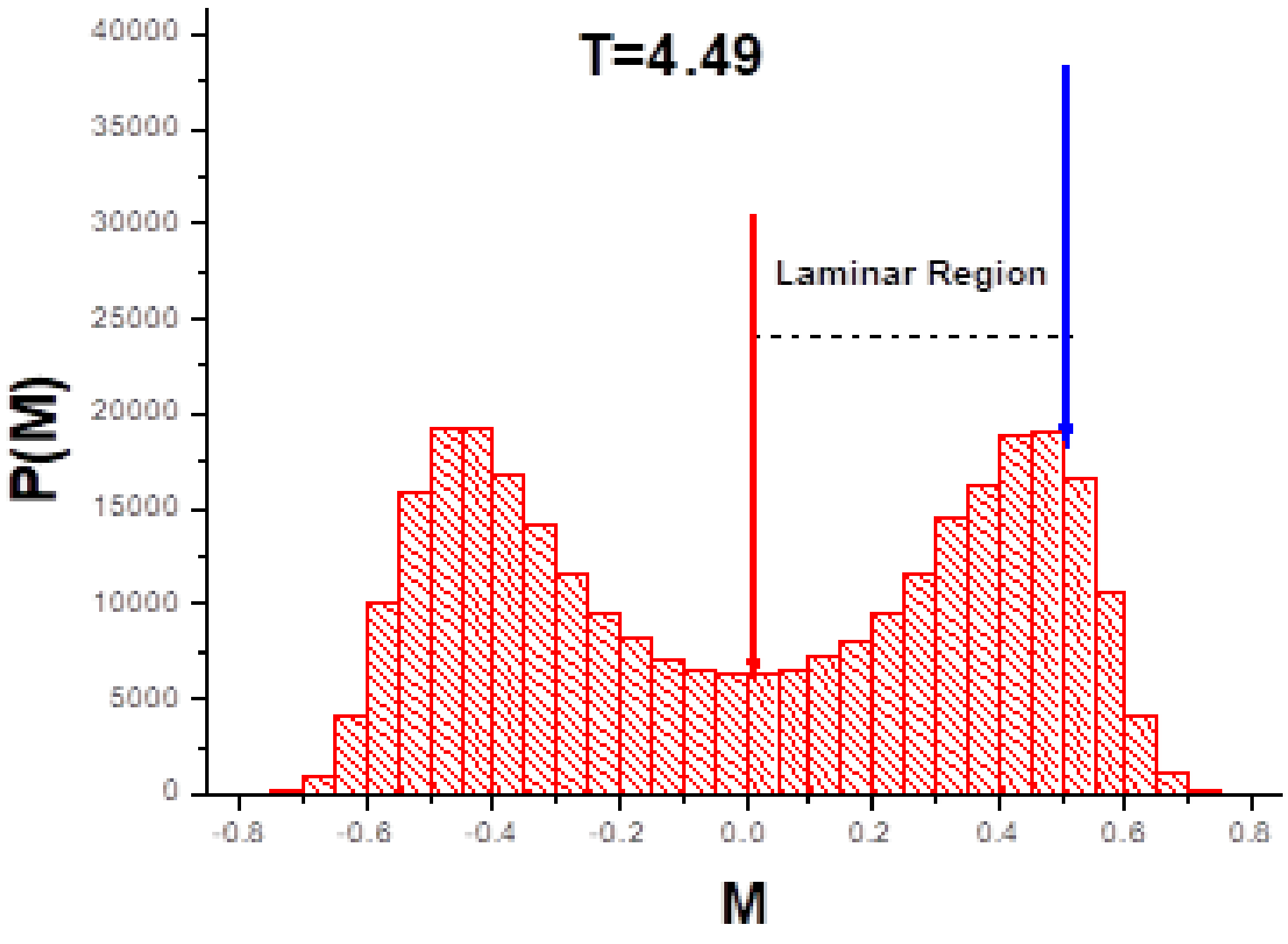

**Fig. 8.** Distribution of magnetization values at temperature $T = 4.49$.

The large width of the laminar region at $T = 4.49$ implies correspondingly large laminar lengths $L$. Therefore, the mean laminar length $\langle L \rangle$ is expected to be maximized at this temperature. Indeed, as shown in Fig. 9, a resonance peak appears in the dependence of $\langle L \rangle$ on temperature.

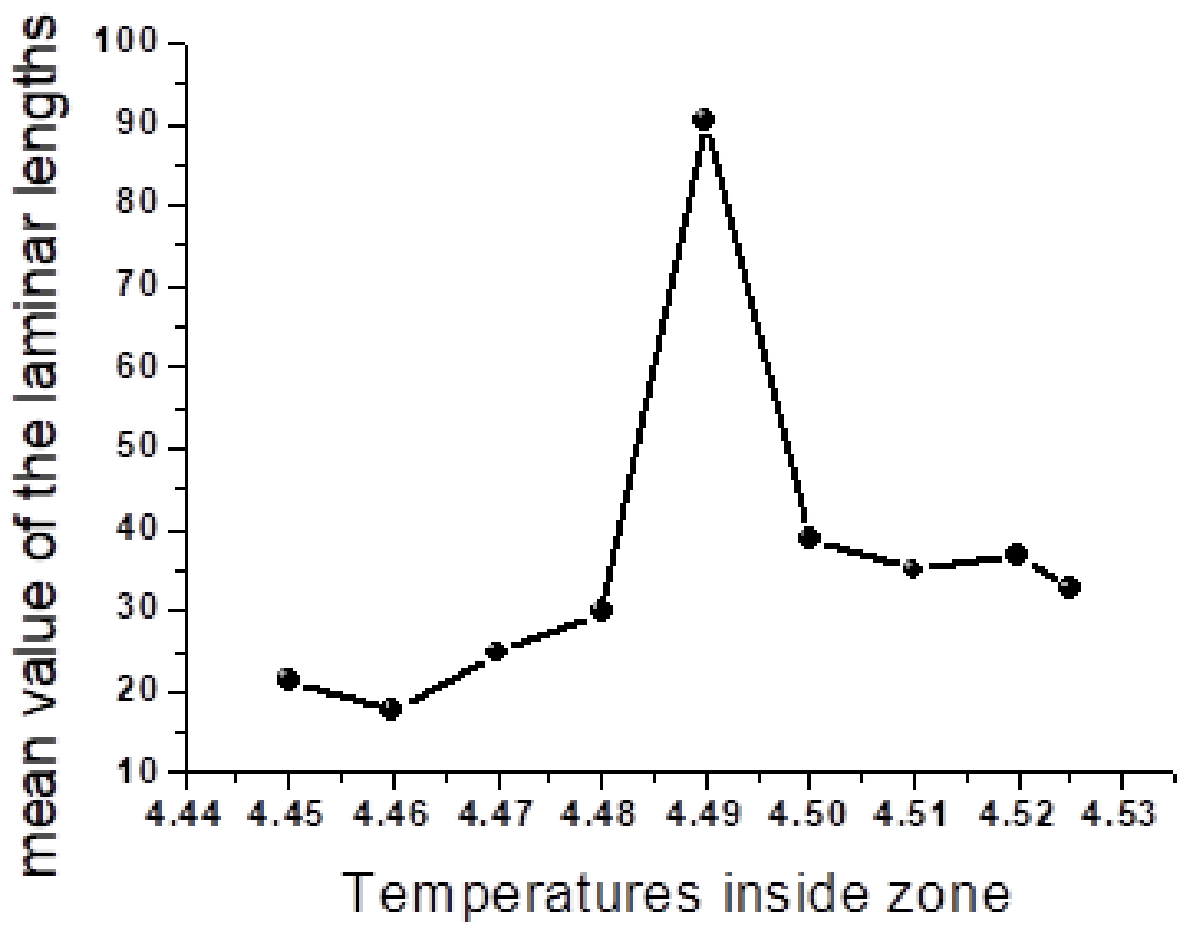

**Fig. 9.** Resonance phenomenon inside the hysteresis zone.

Such a resonance has already been presented in a previous work [23]. Specifically, it was demonstrated that continuous phase transitions exhibit a resonance phenomenon in the temporal fluctuations of the order parameter, occurring whenever the system enters its critical region. This resonance is manifested in the mean value of the laminar lengths (waiting times), $\langle L \rangle$, of the order parameter time-series. By varying the control parameter within the critical region, we found that $\langle L \rangle$ becomes maximum at the (pseudo)critical point. In [23], this was demonstrated for numerical data of the 3D-Ising model, the control parameter of which is the temperature, as well as using experimental data of the voltage across the resistor of a resistor– inductor–diode (RLD) electronic circuit, the control parameter of which is the driving signal frequency.

For the 3D-Ising model studied here, the resonance also manifests in Fig. 9 as a maximum in the mean laminar length, $\langle L \rangle$, but within the hysteresis zone of this finite-size size system. The resonance of Fig. 9 occurs at $T = 4.49$, corresponding to the midpoint of the hysteresis interval $\Delta\mathrm{T}$. It results from the competition between critical dynamics associated with the unstable fixed point and non-critical dynamics imposed by the emerging vacuum. The shape of the resonance resembles that of a $\lambda-$transition [24], characteristic of second-order phase transitions in superfluids and superconductors. This analogy is valid because the Fig. 9 represents a phase-space plot: the vertical axis corresponds to laminar lengths generated by magnetization fluctuations, and the horizontal axis represents the control parameter (temperature).

Beyond the above-mentioned physical interpretation of the observed resonance, a mathematical interpretation can also be given investigating Eq. (4). Equation (4) provides a mathematical description of intermittency, which differs from real physical systems due to the presence of stochastic noise. Nevertheless, as shown in previous studies, the dynamics of real systems at criticality coincide with those of mathematical intermittency [10].

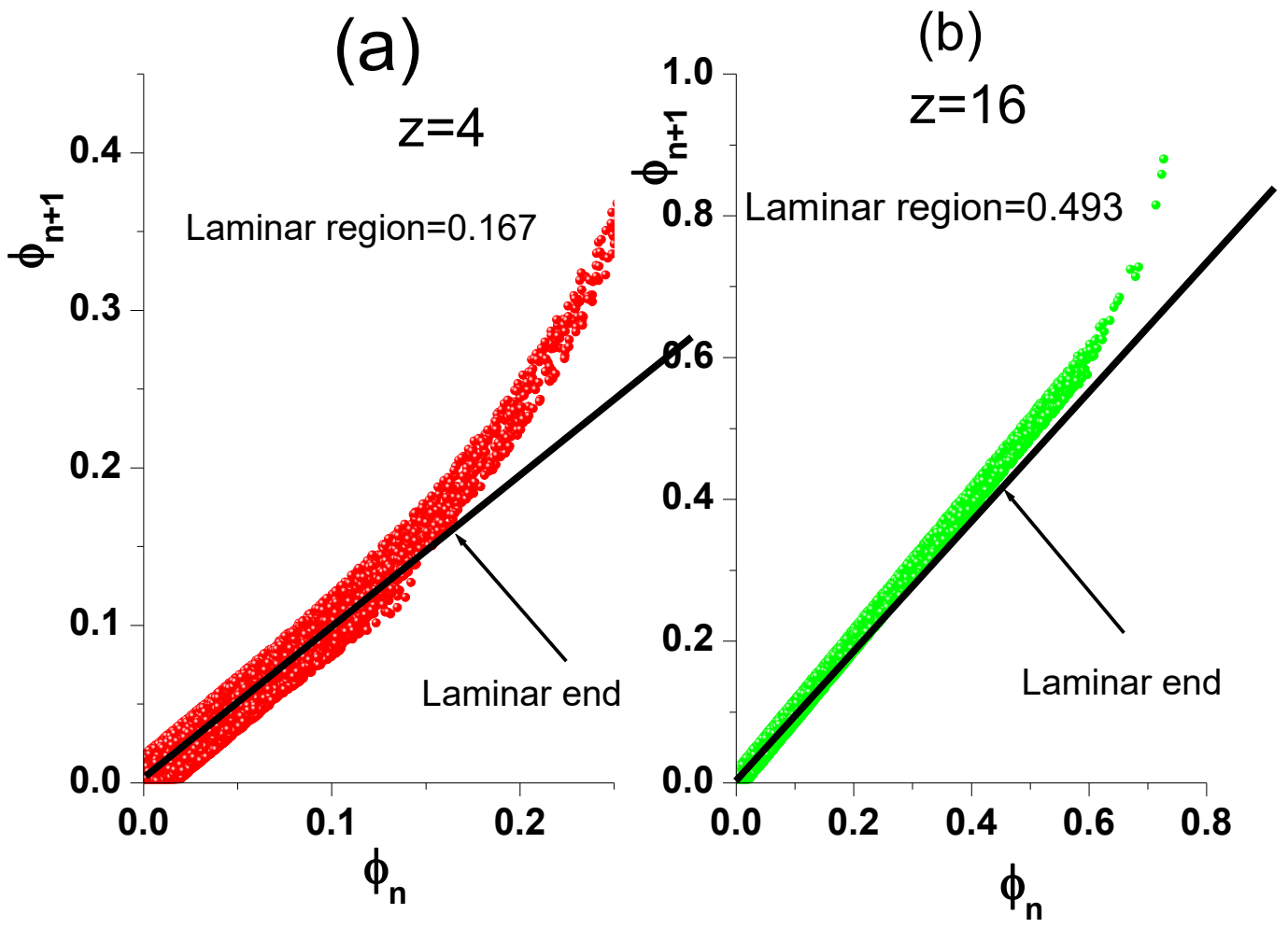

**Fig. 10.** (a) Return map for type-I intermittency (Eq. (4)) with $z = 4$. The laminar region extends from $0$ to approximately $0.167$. (b) Same map for $z = 16$, with all other parameters unchanged. The laminar region now extends to approximately $0.493(> 0.167)$.

Let's focus on the nonlinear term in Eq. (4), $u \cdot \phi_n^z$. As the order parameter increases, for fixed $u$ and $z$, this nonlinear term grows, increasing the distance from the bisector in the return map. Once this distance exceeds a threshold, the laminar phase terminates and chaotic bursts emerge. Equation (4) contains three parameters. In Fig. 10, we investigate the effect of varying only the exponent $z$, while keeping the coupling constant $u$ and noise width $\epsilon$ fixed. It is evident from Fig. 10 that as $z$ increases, the laminar region expands, implying an increase in the mean laminar length. So, we see that one way to increase in the value of laminar ends and therefore the maximalization of the mean value of laminar lengths is the increase in the value of the exponent of the non-linear term of Eq. (4). However, is this the mechanism responsible for the observed resonance?

We remind that for $T = 4.49$, which corresponds to resonance, the lowest exponent $p \in [1,2)$ –near unity– was obtained among the studied temperatures within the hysteresis zone. Interestingly, from Eq. (5) and (7), derived from intermittency theory, we obtain:

$$p = \frac{z}{z-1} = \frac{1}{1-\frac{1}{z}}. \tag{14}$$

Taking the limit $z \to \infty$ we find $p \to 1^{+}$. Thus, the value $p = 1.03$ obtained for $T = 4.49$, corresponding to the observed resonance, is consistent with intermittency theory.

Another way to strengthen the nonlinear term, would be to keep the exponent $z$ constant and highly increase the coupling constant $u$. But in such a case there is no mathematical relation between the parameter $u$ and the exponent $p$ in analogy to Eq. (14). Thus, we cannot explain the obtained results through mathematical intermittency in such a case. However, it is certain that a large value of $u$ would assist the lengths of the laminar regions to grow, so the resonance will be clearer.

Summarizing the up to now findings, one could conclude that for finite-size systems the second order phase transition according to $\phi^4$ Landau theory takes place through a resonance phenomenon.

As already mentioned in Section 3, as the system gets smaller, the hysteresis zone gets wider. So, the corresponding resonance would have wider width.

## 7. Decay of the Resonance in the Euclidean Space

Tachyons are particles with imaginary mass and superluminal velocity [1,2]. In infinite-size systems, $\phi^4$ theory predicts tachyons to be unstable and unable to exist in real time [2,25,26]. In finite-size systems, however, tachyons persist within the hysteresis zone up to the SSB point, but they continue to violate the causality principle [1,27]. As shown in [1], the appropriate framework for their existence is the Euclidean space, where time is imaginary.

The soliton mass at SSB is given by [2]:

$$M_s \sim bm^3, \tag{15}$$

where $b$ is the $\phi^4$ coupling constant (Eq. (1)) and $m$ is the tachyon mass. Since $m$ is imaginary, the soliton mass ($\sim m^3$) is also imaginary. As already mentioned in the Introduction, the resonances in particle physics are in many cases excitations that decay into particles (or quasi particles). Because in the hysteresis zone the tachyons and solitons "live" in the Euclidian space, we will attempt to transfer the resonance of Fig. 9 in the Euclidian space. This way, we could investigate a possible relation between resonance and imaginary particles.

The relation between real time $t$ and imaginary time $\tau$ is [28]:

$$t = -\mathrm{i}\tau. \tag{16}$$

From the Feynman path-integral formulation [28], one obtains:

$$\tau = \frac{1}{kT}. \tag{17}$$

Setting $k = 1$ yields:

$$\tau = \frac{1}{T}. \tag{18}$$

Applying this transformation to Fig. 9 produces Fig. 11, which resembles a biological spike (Fig. 12).

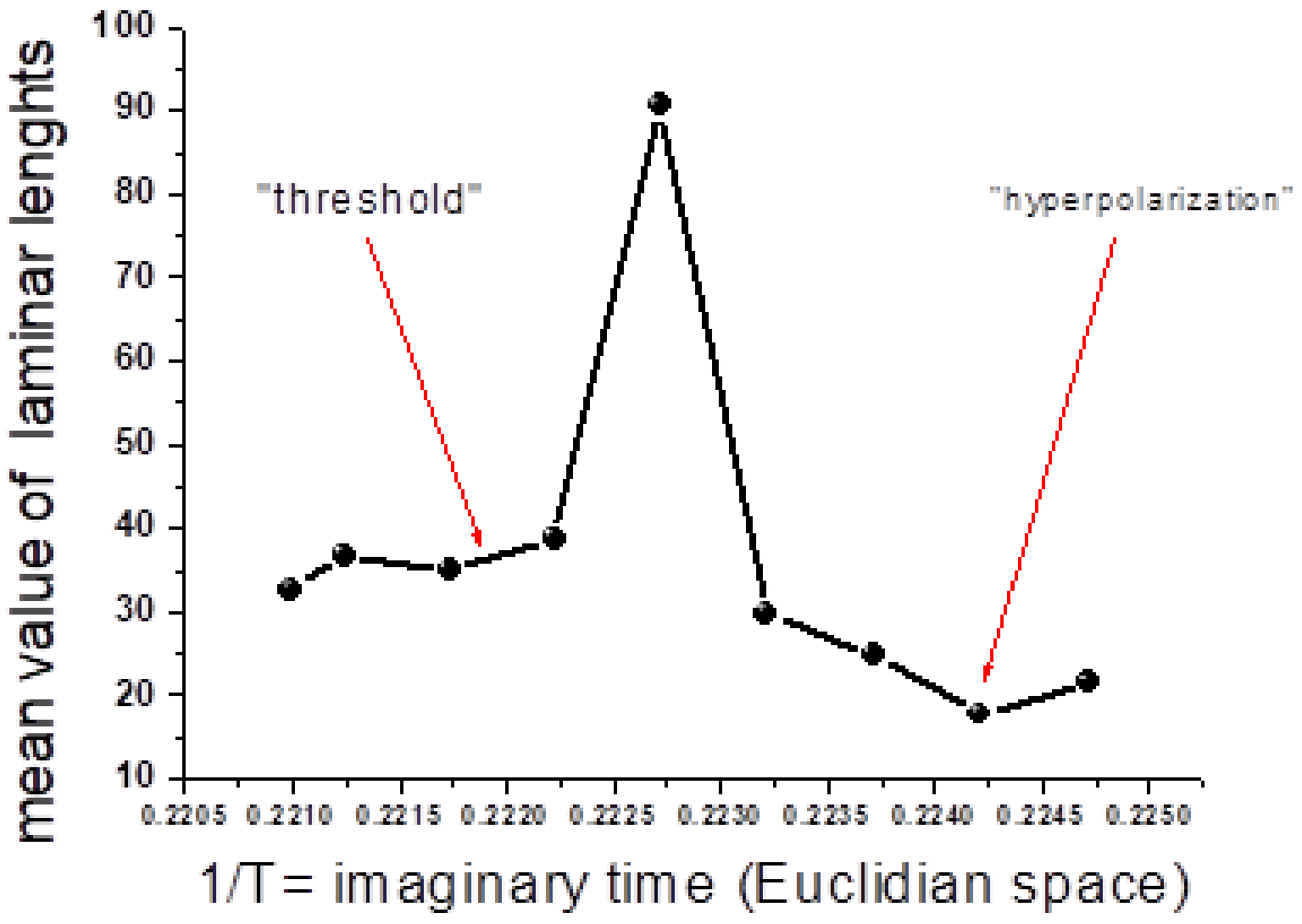

**Fig. 11.** Resonance in the Euclidean space, obtained by inversion of the temperature axis. The red-arrow indicated segments resemble characteristic phases of a biological spike (Fig. 12).

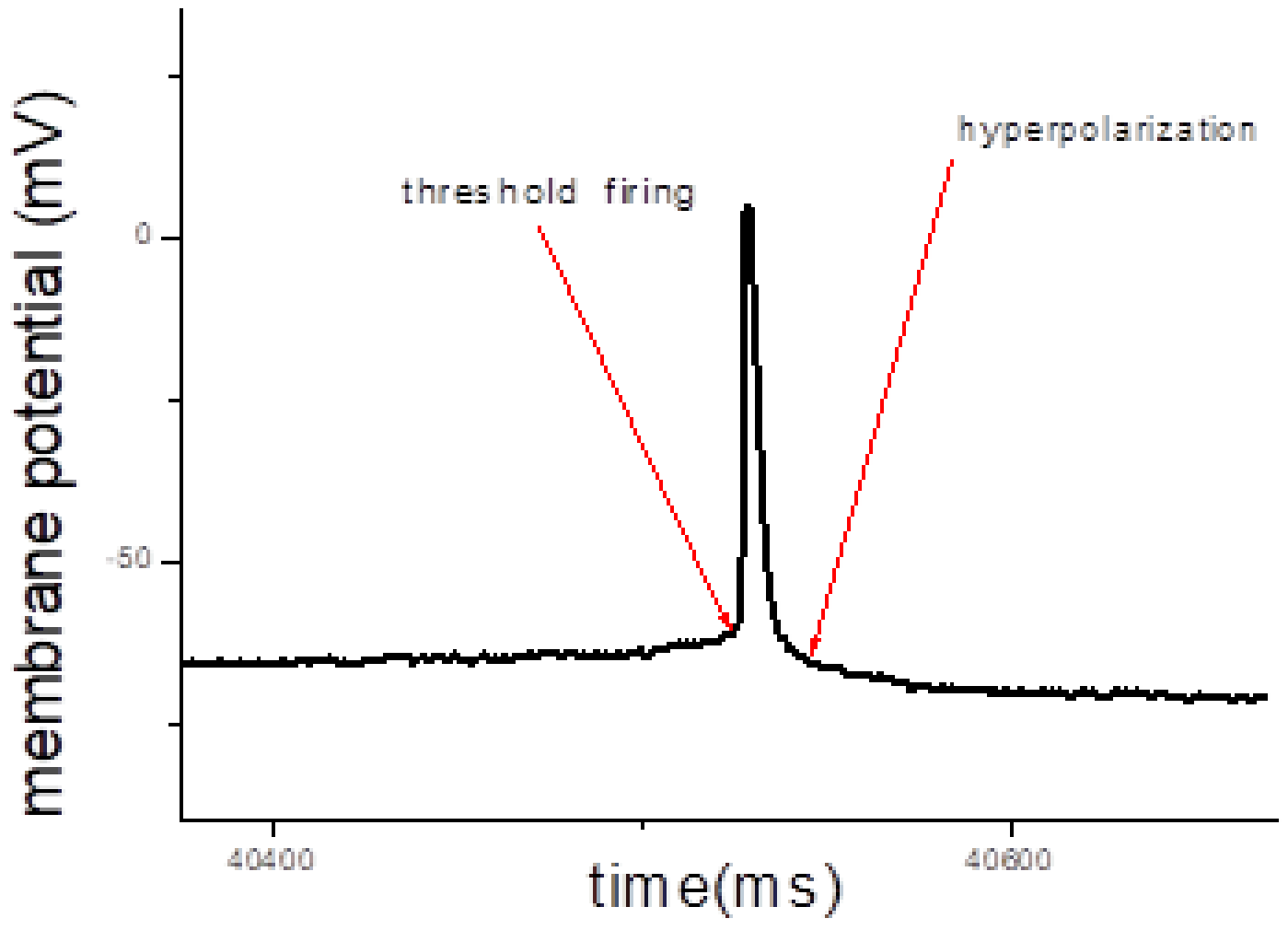

**Fig. 12.** A biological spike in the real time.

**8. Classical-to-Quantum Transition through Resonance?**

We suggest that the phenomena observed within the hysteresis zone represent a transition from the classical regime to a quantum-like regime governed by thermal fluctuations. The initial classical critical state (see Fig. 2) is characterized by positive and negative magnetization values fluctuating symmetrically around zero. As the control parameter, namely the temperature, decreases below its pseudocritical value in a finite system, the emergence of two distinct lobes corresponding to the positive and negative orientations of the classical magnetic moments (spins) begins.

However, inspection of the magnetization distributions (Fig. 3) for the initial temperatures in the range $T = 4.525$ down to $T = 4.50$ reveals that the two lobes remain strongly interconnected. This indicates that the system is still in a classical state. This conclusion is further supported by the fact that the exponents of the laminar-length distributions, which reflect the system dynamics, remain equal to the corresponding exponent of the critical state. Thus, up to $T = 4.50$, the system continues to behave classically due to a classical entanglement between the two lobes. This occurs because, at relatively high temperatures, thermal fluctuations are strong enough to allow communication between the two lobes.

After the resonance at $T = 4.49$ takes place, the system behavior changes qualitatively. As the temperature decreases further, the two lobes gradually separate, reaching almost complete detachment at the spontaneous symmetry breaking (SSB) temperature. At this point, two discrete states emerge, $M_{\uparrow}, M_{\downarrow}$.

The simultaneous presence of these two distinct states at SSB is a hallmark of superposition, a concept that belongs to quantum theory. It must be emphasized, however, that this superposition does not involve pure quantum states, but rather the distributions of thermal fluctuations of the magnetization field. In the next section we will discuss this topic in detail.

This superposition can be seen from the fact that, when the system is cooled below the SSB temperature, only one of the two states is observed, either positive or negative magnetization fluctuations, thereby restoring classical behavior. Which of the two classical states is realized depends on the initial configuration of the 3D-Ising system, namely the initial spin arrangement on the lattice.

Thus, the previously unidentified resonance appearing near the center of the hysteresis zone delineates the boundary between a classical regime, where the two states communicate, and a macroscopic "quantum" regime, where the two states coexist in superposition at SSB.

What does this imply? If one traverses the hysteresis zone from high to low temperatures, upon reaching the resonance region one begins to observe the coexistence of two states simultaneously. At the SSB point, these states become clearly separated. This scenario challenges the prevailing view that observation necessarily destroys quantum superposition through wave-function collapse, rendering it unobservable. Since the width of the hysteresis zone increases as the system size decreases, it follows that in microscopic systems, such as those involving elementary particles, this zone may be sufficiently wide to allow direct observation of quantum behavior without collapse.

It is important to reiterate that the "quantum world" discussed here does not refer to discrete quantum numbers, but to quantum thermal fluctuations, manifested as two separated lobes in the magnetization distribution. A clear distinction must be drawn between quantum vacuum fluctuations and thermal fluctuations of a quantum field. In the present work, we focus exclusively on the latter, namely thermal fluctuations of the magnetization field, which acquire quantum-like properties such as superposition in a narrow region of the hysteresis zone near SSB.

We emphasize that the phase transition examined in this study is described by a classical theoretical framework and refers to a macroscopic classical system, namely the 3D Ising model. Nevertheless, nature provides several examples of macroscopic systems that exhibit quantum fluctuations. In quantum field theory, the Casimir effect [30,31] is a force acting between macroscopic boundaries due to quantum vacuum fluctuations. In July 2020, it was reported that quantum vacuum fluctuations can influence the motion of macroscopic, human-scale objects, as evidenced by measurements of correlations below the standard quantum limit in the position–momentum uncertainty of the mirrors of LIGO [32,33]. Furthermore, in 1962, B. D. Josephson predicted quantum tunneling effects in macroscopic superconducting systems [34,35].

## 9. Uncertainty Principle inside the Hysteresis Zone

In quantum mechanics, the uncertainty relation reads:

$$\Delta p \cdot \Delta x \geq \frac{\hbar}{2}, \tag{19}$$

where $h$ is the Planck constant.

Writing $\Delta p = M \cdot \Delta u$, where $M$ is the mass of virtual particle (or Gauge fields –such as electric and magnetic fields which represent the electromagnetic force carried by photons, W and Z fields which carry the weak force, and gluon fields which carry the strong force), gives:

$$\Delta u \cdot \Delta x \geq \frac{\hbar}{2M}. \qquad (20)$$

In phase space, the product $\Delta u \cdot \Delta x = \Delta\Omega$ represents the area of the surface defined by the set of points in phase space. An analogous quantity can be defined for the generalized phase-space diagram of the magnetization field, where $\Delta\Omega$ characterizes the phase-space area occupied by the magnetization dynamics. Thus from Eq. (20), one obtains:

$$K \leq 2M \cdot \Delta\Omega. \qquad (21)$$

Here, $K$ denotes a new constant that plays a role analogous to the Planck constant in the generalized phase space. For the quantity $K$ to remain constant, the product appearing in relation (21) must also remain constant. As $\Delta\Omega$ becomes smaller, $K$ correspondingly decreases. Conversely, in order to preserve $K$ as a constant, $\Delta\Omega$ must attain its minimum possible value within the hysteresis zone, which occurs precisely at the spontaneous symmetry breaking (SSB) point. In Fig. 13, we confirm that such behavior indeed takes place.

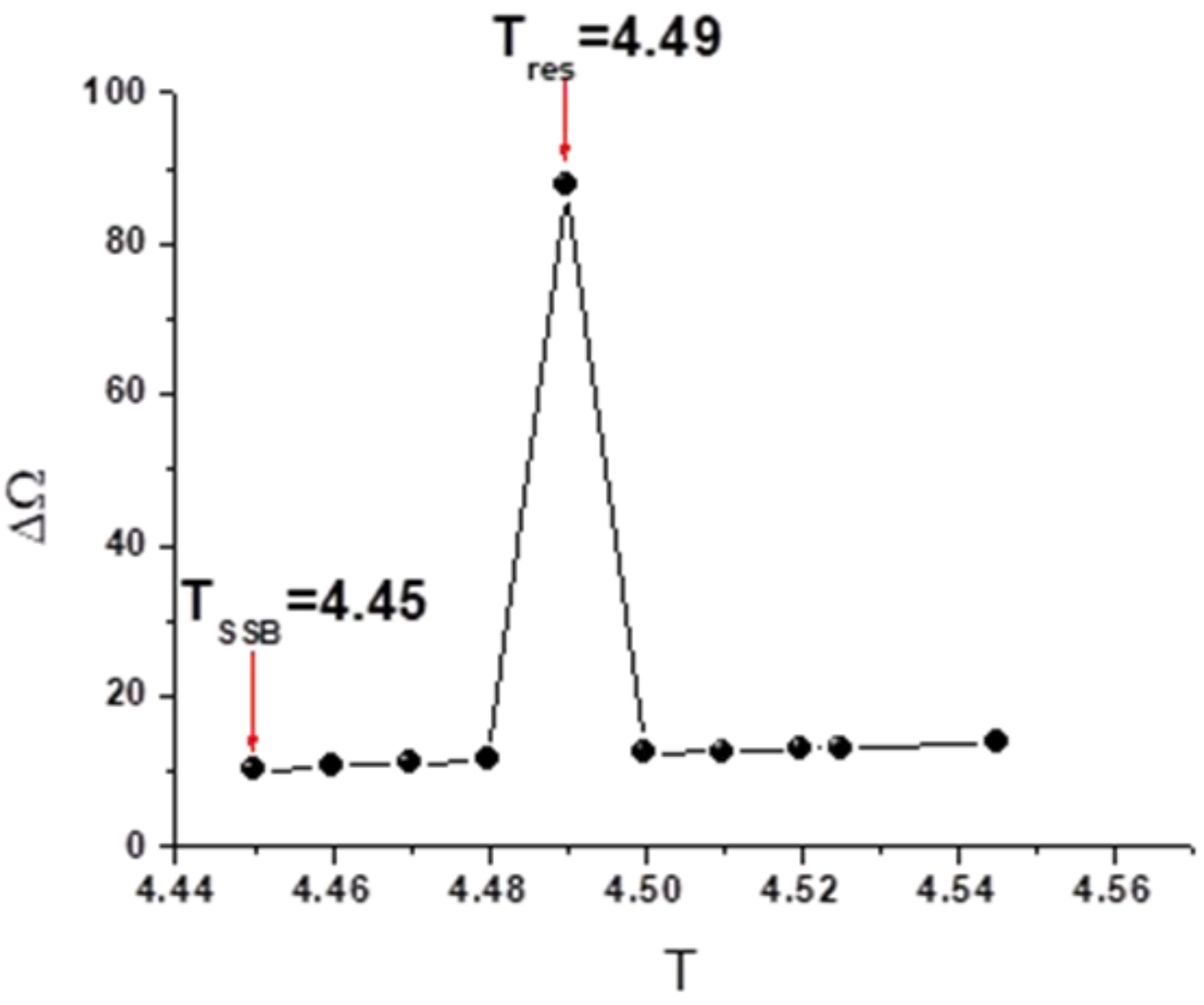

**Fig. 13.** Plot of $\Delta\Omega$ in phase space as a function of temperature within the hysteresis zone. The decrease of $\Delta\Omega$ with decreasing temperature is monotonic, and at the final point of the zone, corresponding to SSB, $\Delta\Omega$ reaches its minimum value. An exception is observed at the special resonance temperature $\mathrm{T} = 4.49$, where anomalous behavior appears.

As the system size decreases, the hysteresis zone broadens and the SSB point shifts toward lower temperatures. This implies that $\Delta\Omega$ shrinks, and consequently the amplitude of quantum thermal fluctuations decreases. When the system dimensions approach the scale of $\hbar$, the quantum thermal fluctuations $\Delta\Omega$ approach discrete quantum dots. In this limit, the constant $K$ tends toward the Planck constant.

In Fig. 14, we present the time series of the thermal fluctuations at the SSB temperature. This time series is very close to exhibiting ideal on–off intermittency. As $K \to \hbar$, the thermal fluctuations degenerate into line-like structures, and time progressively loses the significance it possesses in classical physics, as is also the case in pure quantum mechanics.

Therefore, when the system enters the dimensional regime of the quantum world ($\sim \hbar$), the classical behavior, represented by the continuous black fluctuations in Fig. 14, degenerates into discrete quantum dots (red lines). This transition occurs only up to the SSB point of a second-order phase transition, such as those believed to have occurred during the earliest moments of the universe, for example during matter–antimatter creation. The positive and negative fluctuations are completely separated. As shown in Ref. [13], the autocorrelation function of the time series at SSB attains its maximum possible value, which in normalized units approaches unity. Consequently, the two magnetization states, $M_{\uparrow}, M_{\downarrow}$, are highly correlated, analogous to a quantum superposition, despite the fact that the 3D Ising model is a macroscopic system.

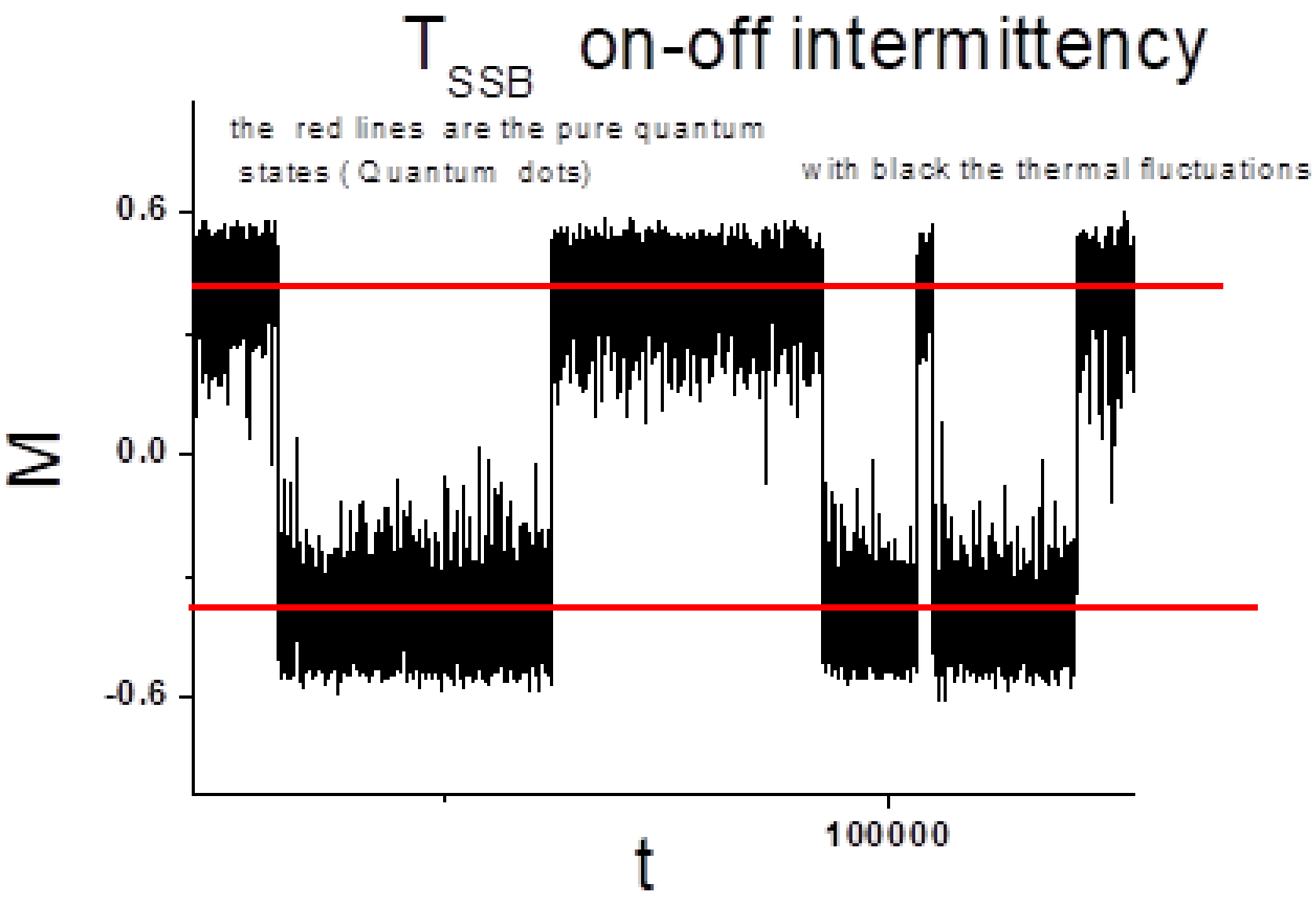

**Fig. 14.** Segment of the time series of thermal fluctuations at the SSB temperature.

As the system size decreases, the hysteresis zone expands and the SSB effectively begins to "freeze" due to the decreasing temperature, while thermal fluctuations collapse into quantum dots (red lines). This process constitutes a complete evolution from a classical system to a quantum system. From the moment that superposition of quantum thermal fluctuations emerges at SSB, the system may therefore be regarded as a macroscopic quantum system.

## 10. Conclusions

In this work, we demonstrated the existence of a resonance within the hysteresis zone, preceding the spontaneous symmetry breaking (SSB) of a second-order phase transition. A key observation is that, when this resonance is analytically continued to Euclidean space, where time becomes imaginary and effectively represents a fourth spatial dimension, its decay products are restricted to the only excitations that can exist within the hysteresis zone, namely tachyons. In Euclidean space, tachyons possess the important property that they do not violate the principle of causality.

Finally, we showed that this resonance delineates the transition from a classical regime to a quantum-like regime in macroscopic thermal systems. Moreover, through the expansion of the hysteresis zone as the system size decreases, this transition may extend even into the microscopic quantum domain.